\documentclass[twocolumn,prd,aps,showpacs,tightenlines,nofootinbib,preprintnumbers,]{revtex4}

\usepackage{graphicx}
\usepackage{dcolumn}
\usepackage{epsfig}
\usepackage{amssymb}
\usepackage{amsmath}

\def\fun#1#2{\lower3.6pt\vbox{\baselineskip0pt\lineskip.9pt
\align{$\mathsurround=0pt#1\hfil##\hfil$\crcr#2\crcr\sim\crcr}}}

\begin{document}

\title{Testing gravity on kiloparsec scales with strong gravitational lenses}

\author{Tristan L. Smith}
\affiliation{Berkeley Center for Cosmological Physics \\ Physics Department, University of California, Berkeley, CA 94720}
\date{\today}
\begin{abstract}
Modifications to GR generically predict time and scale-dependent effects which may be probed by observations of strong lensing by galaxies.  Measurements of the stellar velocity dispersion determine the dynamical mass whereas measurements of the Einstein radius determine the lensing mass.  In GR these two masses are equal; in alternative gravity theories they may not be.  Using measurements of the stellar velocity dispersion and strong lensing around galaxies from the Sloan Lens ACS (SLACS) survey we place constraints on lensing in modified gravity theories and extend previous studies by applying this data to explore its dependence on various properties of the lens such as the lens redshift or mass and thereby constrain scalar-tensor, $f(R)$ gravity theories, and generic parameterizations of deviations from GR.  Besides applying the observations to these specific gravity theories, the data places a constraint on a generic dependence of modifications to GR on the lens mass and redshift.  At the 68\% confidence level we find that the ratio between the lensing and dynamical masses can only vary by less then $50\%$ over a mass range for the lens galaxies of $10^{12} \lesssim M/M_{\odot} \lesssim 10^{14}$ and less than $40\%$ over the redshift range $0.06 < z < 0.36$. 
\end{abstract}

   \maketitle

\section{Introduction}

The ability to test our basic understanding of gravity has been surprisingly limited \cite{Will:1981cz}.  Most precision tests have concentrated on the motion of the planets and light within the solar system or the motion of binary pulsars.  Although measurements in the solar system have reached the level of testing deviations from general relativity (GR) to one part in $10^5$ \cite{Bertotti:2003rm, shapiro}, they only constrain theories in the weak gravity limit, on scales of an AU ($r \sim 10^{12}$ cm), and at a single redshift, $z=0$.   Binary pulsar systems similarly test theories at $z=0$ and on relatively small scales but have the added aspect of testing gravity in the limit where it is large due to the compact nature of neutron stars \cite{Will:1981cz}.  

The discovery of an accelerated cosmic expansion
\cite{Perlmutter:1998np,Riess:1998cb}  has led to a
flurry of theoretical activity.  Although it is possible to explain the accelerated expansion within GR by introducing a cosmological constant or new cosmic scalar field or other source of energy density, another approach is to consider these observations as the first observational indication of a need for modifications to Einstein's theory of GR \cite{Caldwell:2009ix}. 

Many groups have attempted to explain the accelerated expansion within alternative theories of gravity.  These observations can be explained as the result of the dynamics of a scalar field within a generalized scalar tensor theory \cite{Boisseau:2000pr, EspositoFarese:2000ij,Perivolaropoulos:2005yv,Gannouji:2006jm}.  Another proposal modifies the Einstein-Hilbert action by the addition of a general function of the Ricci scalar, $f(R)$ \cite{Carroll:2003wy,Capozziello:2003tk}.  Depending on the functional form of the function $f(R)$, such a term may give rise to late-time accelerated expansion.  Refs.~\cite{Dvali:2000hr, Deffayet:2001pu},  proposed a five-dimensional theory of gravity which may lead to an epoch of late time accelerated expansion.  In addition to studies dedicated to the observational consequences of specific modified gravity theories there has also been interest in parameterizing generic deviations from GR on cosmological scales \cite{Hu:2007pj,Caldwell:2007cw,Amin:2007wi,Bertschinger:2008zb}.  
 
Many aspects of these modifications can be constrained or even ruled out by considering tests of gravity made in the solar system or through observations of binary pulsars \cite{Wagoner:1970vr,Damour:1996ke,Erickcek:2006vf,Chiba:2006jp}.  However, given that these theories are naturally dynamical and scale dependent, solar system and pulsar tests can be of limited use.  Therefore it is important to test gravity at a variety of scales and redshifts.  In particular, since those modified gravity theories which are able to produce an epoch of late-time accelerated expansion must become dynamically important when the acceleration starts to dominate (around $z\sim 0.5$ \cite{Turner:2001mx}) probes of modifications to GR around these redshifts are of the greatest interest.  

Observations of strong lensing around galaxies present an important and unique opportunity to probe modifications to GR over a range of redshift and on kpc scales. 
The idea to use strong galaxy lenses to constrain modifications to GR was first proposed in Ref.~\cite{sirousse}.  Following this, Ref.~\cite{Bolton:2006yz} was the first to use a data set of 15 strong lenses from the Sloan Lens ACS survey (SLACS) \cite{2005ApJ...624L..21B, 2006ApJ...638..703B}.  The main difference between the analysis presented here and the one in Ref.~\cite{Bolton:2006yz} is that here the analysis is extended beyond constraining a universal value for $\gamma_{\rm PPN}$ and uses the observations to place constraints on how $\gamma_{\rm PPN}$ may depend on various properties of the lens, such as its redshift or mass.  This paper also extends the analysis to the full SLACS data set of 53 lens systems as well as uses a more realistic model for the luminosity profile of the lens galaxy (a Hernquist profile, as opposed to a power-law profile used in Ref.~\cite{Bolton:2006yz}).  The use of a more realistic luminosity profile leads to a significant shift in the best fit $\gamma_{\rm PPN}$. 

This paper is organized as follows.  In Sec.~\ref{basic lensing} we present how scalar modifications to gravity affect the dynamics of massive test particles (i.e., stars) and photon trajectories differently.  We discuss how a comparison between the dynamics and lensing signal leads to a test of gravity.  In Sec.~\ref{ST} we discuss the predictions from general scalar tensor theories with a massive scalar field.  In Sec.~\ref{sec:fR} we discuss the predictions from $f(R)$ gravity and emphasize its ability to rapidly suppress any deviations from GR and how this transition presents unique observational signatures depending on the mass of the lens galaxy.  In Sec.~\ref{measurements} we discuss how measurements of strong galaxy lenses can be used to constrain modified gravity theories.  We present our conclusions in Sec.~\ref{conclusions}.

\section{Lensing and dynamics in weak-field limit of modified gravity theories \label{basic lensing}}

One of the basic ways that we can distinguish between different theories of gravity is through a comparison
between the predicted and observed motion of test particles.  Such tests compare the motion of 
photons (which move on null geodesics) and the motion of non-relativistic massive particles (which move on time-like geodesics).  The different types of geodesics are sensitive to different components of the metric and hence their comparison allows us to measure those components.  

We start with the line element corresponding to weak gravity, 
\begin{equation}
ds^2 = -(1+2 \Psi) dt^2 + (1-2 \Phi) \delta_{ij}dx^i dx^j,
\label{eq:scalar_metric}
\end{equation}
which has been written in the conformal Newtonian gauge and has introduced the two Newtonian potentials, $\Psi$ and $\Phi$. 
The `bare' Newtonian potential is given by the usual Poisson equation, 
 \begin{equation}
\nabla^2 \Phi_N(\vec{x}) = 4 \pi G \rho.
  \end{equation}
A general modification to gravity introduces two new equations.  One relates the potentials to the underlying mass density, 
\begin{equation}
\frac{1}{2}\nabla^2 \left(\Phi + \Psi\right) = 4 \pi \mu G \rho.
\label{eq:mu}
\end{equation}
The other relates the potentials to one another
\begin{equation}
\frac{\Phi}{\Psi} = \gamma_{\rm PPN}.
\label{normal_gamma1}
\end{equation}
Note that GR is regained when $\mu = 1$ and $\gamma_{\rm PPN} = 0$.  In general both $\mu$ and $\gamma_{\rm PPN}$ can depend on a variety of quantities that determine the space-time such as the local mass density, position, redshift, and so forth.  We will see specific examples of this in the following sections. 
  
In order to distinguish between the two Newtonian potentials we compare the dynamics of stars within a galactic halo and the deflection of light around the halo.  The deflection of the image of a background source through an angle $\hat{\alpha}$ is given by 
 \begin{eqnarray}
 \hat{\alpha} &=& \int \vec{\nabla}_{\perp} (\Psi + \Phi) d\ell, \\
 &=& 2 \mu \int \vec{\nabla}_{\perp} \Phi_{N}\ d\ell,
 \label{eq:deflection}
 \end{eqnarray}
 where $\vec{\nabla}_{\perp}$ is the gradient transverse to the photon's unperturbed trajectory and $d\ell$ is a length element along that trajectory and we have assumed that $\mu$ is independent of position on the relevant scales.  
In terms of the bare potential, $\Phi_N$, observations of stellar dynamics through the spherical Jeans equation  \cite{binney} measure the combination
\begin{equation}
\Psi = \frac{2 \mu}{1+\gamma_{\rm PPN}} \Phi_N.
\label{eq:dynamics}
\end{equation}
Therefore with a knowledge of $\Phi_N$ Eqs.~(\ref{eq:deflection}) and (\ref{eq:dynamics}) show that a comparison between lensing and stellar dynamics provides a measurement of $\gamma_{\rm PPN}$.  

In Appendix \ref{basic lensing app} we derive how a general scalar modification to the GR field equations leads to a modified relationship between the lensing and dynamical masses.  The effects of these modifications can be compactly written in terms of an effective source of stress energy that we denote $T_{\rm eff}$ defined in Eq.~(\ref{eq:Teff}).  

\section{Gravitational lensing in general scalar-tensor theories \label{ST}}

Scalar-tensor theories of gravity \cite{1949Natur.164..637J, Brans:1961sx, Bergmann:1968ve, 1970ApJ...161.1059N,Wagoner:1970vr} present us with an example of a class of modified gravity theories that naturally predict a redshift-dependent $\gamma_{\rm PPN}$. 

We will consider a general scalar-tensor theory defined by the action 
\begin{equation}
S = \frac{1}{2 \kappa} \int d^4 x \sqrt{-g} \left[ \varphi R - \frac{\omega_{\rm BD}(\varphi)}{\varphi} (\partial_{\alpha} \varphi)^2 -2 U(\varphi)\right] + S_m.
\end{equation}
The gravitational field equation is given by
\begin{eqnarray}
\varphi G_{\mu \nu} &+& g_{\mu \nu} \left[ \frac{1}{2}\left(\partial_{\alpha} \varphi\right)^2+ \Box \varphi + U(\varphi)\right] \nonumber \\&-& \frac{\omega_{\rm BD}(\varphi)}{\varphi} \partial_{\mu} \varphi \partial_{\nu} \varphi - \nabla_{\mu} \nabla_{\nu} \varphi = \kappa T_{\mu \nu},
\end{eqnarray}
with the scalar field equation given by
\begin{equation}
\frac{2 \omega_{\rm BD}(\varphi)}{\varphi} \Box \varphi = - R - \left(\frac{\omega_{\rm BD}'}{\varphi} - \frac{\omega_{\rm BD}}{\varphi ^2}\right) (\partial_{\alpha} \varphi)^2 + 2 U', 
\end{equation}
where a prime denotes differentiation with respect to the field $\varphi$.  

Linearizing the scalar field around its cosmological value, $\varphi = \varphi_0 + \delta \varphi $, and specializing to a static case we have
\begin{eqnarray}
T_{\rm eff} &=& \nabla^2 \delta \varphi + 2 U' \delta \varphi,\label{eq:ST_Teff}\\
\nabla^2 \delta \varphi &=& \frac{\kappa T}{3+ 2 \omega_{\rm BD}(\varphi_0)} + m^2 \delta \varphi,
\label{eq:ST_scalar}
\end{eqnarray}
where the scalar field mass, $m$, is a function of the background field and derivatives of the potential $U$ and Brans-Dicke function, $\omega_{\rm BD}$, whose exact form is not needed for this discussion.  In general scalar-tensor models which produce late-time acceleration have both $m \sim H$ and $U \sim H$ so that the mass and potential are negligible on kpc scales.  
In the absence of a potential and scalar field mass we then find the standard result \cite{Brans:1961sx}
\begin{equation}
\frac{1+\gamma_{\rm PPN} }{2} = \frac{3 + 2\omega_{\rm BD} [\varphi_0(z)]}{4 + 2\omega_{\rm BD}[\varphi_0(z)]},
\end{equation}
and $\gamma_{\rm PPN}$ then depends on time through the background evolution of $\varphi_0$.  

Soon after the expansion of the universe was shown to be accelerating many groups proposed scalar-tensor models as an explanation.  In order to produce models with expansion histories in agreement with observations Refs.~\cite{Boisseau:2000pr,Gannouji:2006jm, EspositoFarese:2000ij,Perivolaropoulos:2005yv} established an algorithm by which observations of both the expansion history as well as the growth of structure would enable a complete determination of the scalar-tensor theory.  In particular, Refs.~\cite{Boisseau:2000pr,EspositoFarese:2000ij} realized that the specification of the expansion history [in the form of $H(z)$] allows a reconstruction of $\varphi(z)$; in the absence of precise measurements of the growth rate a functional form for $U(\varphi)$ must be specified from which follows $\gamma_{\rm PPN}(z)$.  

An interesting case considers the ability to produce accelerated expansion by introducing a cosmological constant, $\Omega_V$, within a scalar-tensor theory which is less than its value required in the standard $\Lambda$CDM cosmology (i.e., $\Omega_{\Lambda} \simeq 0.7$).  As discussed in Ref.~\cite{EspositoFarese:2000ij} these models remain viable until some $z_{\rm max}$ (which is typically of order unity) after which the theory becomes inconsistent\footnote{As is discussed in detail in Ref.~\cite{EspositoFarese:2000ij} for $z > z_{\rm max}$ the graviton carries negative energy.}.  Therefore, scalar-tensor theories with a cosmological constant $\Omega_V < \Omega_{\Lambda}$ can only explain the observed accelerated expansion up to $z_{\rm max}$.  This implies that an observation of the expansion history at redshifts greater than $z_{\rm max}$ can rule these models out.  Because of this Ref.~\cite{EspositoFarese:2000ij} emphasizes that, as opposed to solar system observations, measurements of the luminosity distance, $D_L(z)$, at 
larger redshifts will place the most stringent constraint on these theories.  Here we shall see that a measurement of $\gamma_{\rm PPN}$ at a range of redshifts less than $z_{\rm max}$ can also serve to distinguish between these models and $\Lambda$CDM. 

\begin{figure}[!h]
\centerline{\epsfig{file=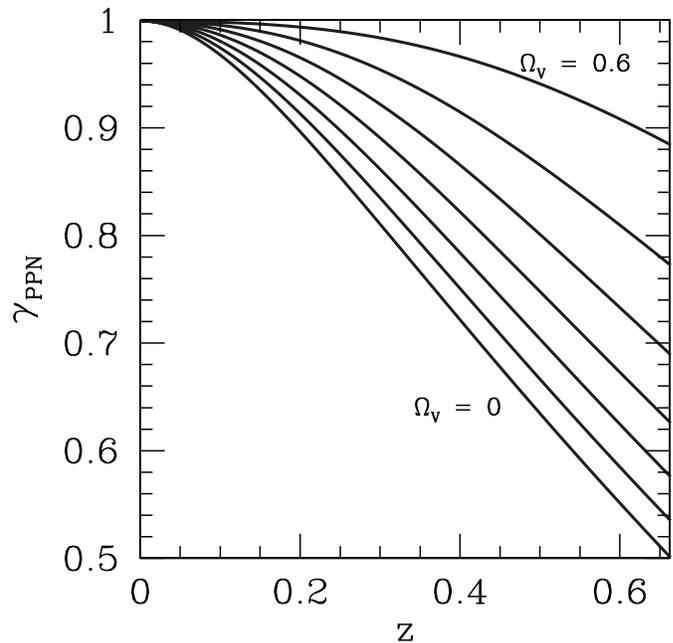,height=20pc,angle=0}}
\caption{The evolution of $\gamma_{\rm PPN}$ in scalar tensor theories which exactly mimic the flat $\Lambda$CDM expansion history with $\Omega_M = 0.3$. The various curves correspond to different choices of $\Omega_V$ going from zero on the bottom to 0.6 at the top in steps of 0.1.  As $\Omega_V$ increases towards the value of $\Omega_{\Lambda} = 0.7$ the differences between the scalar tensor theory and GR decreases.  For all cases presented here the parameters of the theory have been chosen to be indistinguishable from general 
relativity at $z=0$ (i.e., in the solar system).}
\label{fig:transition}
\end{figure}

The choice of $\Omega_V$ in scalar-tensor theories fully specifies the potential $U$ which then allows a calculation of $\gamma_{\rm PPN}(z)$.  We show the evolution of $\gamma_{\rm PPN}$ for various choices of $\Omega_V$ in Fig.~\ref{fig:transition}.  Note that for the scalar-tensor models considered here the parameters have been chosen so as to be indistinguishable from GR at $z=0$ (i.e., in the solar system).  Also note that as $\Omega_{V}$ approaches the $\Lambda$CDM value of 0.7 we regain GR at all redshifts and $\gamma_{\rm PPN} = 1$.  

\section{Gravitational lensing in $f(R)$ gravity \label{sec:fR}}

A sub-class of general scalar-tensor theories that present markedly unique predictions are those theories for which the Brans-Dicke parameter identically vanishes.  Recently, these theories have been extensively studied in particular case of $f(R)$-theories \cite{Carroll:2003wy,Capozziello:2003tk}.  These theories contain a particularly interesting mechanism, known as the chameleon mechanism \cite{Khoury:2003aq, Khoury:2003rn}, in which the modifications to GR are rapidly suppressed around an object with sufficient density. This rapid change in the behavior of the theory presents a unique scale-dependent lensing signature.  

The action for $f(R)$-theories takes the form
\begin{equation}
S=\frac{1}{2\kappa} \int d^4x \sqrt{-g}\left[R+f(R)\right]+S_m,
\label{action}
\end{equation}
where $f(R)$ is a function of the Ricci scalar $R$ and $S_m$ is the matter action. 
The gravitational field equation can be written as
\begin{eqnarray}
\left[1+f'(R)\right]G_{\mu \nu}\nonumber &+& \frac{1}{2} g_{\mu \nu} \left[R f'(R) - f+2 \Box f'(R)\right] \nonumber \\&&- \nabla_{\mu} \nabla_{\mu} f'(R)= \kappa T_{\mu \nu}.
\label{eq:fR_fieldequation}
\end{eqnarray}
In this theory the Ricci scalar becomes a dynamical quantity whose equation of motion is determined by the trace of the field equation, 
\begin{equation}
\Box f'(R) = \frac{1}{3} \bigg(\kappa T + R\left[1-f'(R)\right]+2f\bigg).
\label{eq:trace}
\end{equation}
In the limit where $f \rightarrow 0$, Eq.~(\ref{eq:trace}) implies the usual algebraic relationship $R = - \kappa T$ and GR is regained and $T$ is the trace of the usual stress-energy tensor. 

Using the trace equation to rewrite the gravitational field equation this theory produces
\begin{equation}
T_{\rm eff} = \frac{1}{3\kappa} \left[\kappa T + R\right] + \frac{1}{3\kappa} \left[ 2 R f'(R) - f\right].
\end{equation}
Solutions to the trace equation, Eq.~(\ref{eq:trace}), determine the lensing predictions for this theory.  To understand these solutions, we rewrite the trace of the field equation as
\begin{equation}
\Box f'(R) + \frac{\mathrm{d} V}{\mathrm{d} f'(R)}=0, 
\end{equation}
with
\begin{equation}
\frac{\mathrm{d} V}{\mathrm{d} f'(R)} \equiv \frac{1}{3} \bigg(\kappa T + R\left[1-f'(R)\right]+2f\bigg).
\end{equation}
We note that for functions $f(R)$ which reproduce the observed expansion history, the minimum of this potential yields the general relativistic relationship between $R$ and $T$, $R= -\kappa T$ \citep{Hu:2007nk}. 

Energetics drive the solution of Eq.~(\ref{eq:trace}) towards two limiting cases \citep{Hu:2007nk}.  If $f'(R)$ remains close to its asymptotic, cosmological, value as we move within the galaxy this trades the energy cost of fixing the $f'(R)$ at a high point in its effective potential against the gain in maintaining a nearly homogeneous field.  In this case, the solution to Eq.~(\ref{eq:trace}) can be found by linearizing $f'(R)$ around its cosmological value and we have $T_{\rm eff} = T/3$ so that $\gamma_{\rm PPN} = 1/2$.  On the other hand, if the scalar curvature is able to reach the minimum of its potential this will be at a cost in gradient energy since the scalar curvature will have to transition from its asymptotic value, $R_0$, to the general relativistic value $R = - \kappa T$.  At the minimum of the potential deviations from GR are highly suppressed.  

Within a given object, far away from the center the scalar curvature starts off near its asymptotic value, $R_0$, and evolves with radial distance from the center.   If the object is too `small', in a sense we will make clear in a moment, then $R \sim R_0$ throughout the object and deviations from relativity will be of order unity.  On the other hand, if the object is compact enough then the scalar curvature is forced to the minimum of its potential and $R = - \kappa T$ within some radius $r_0$.  Within that radius deviations from relativity are highly suppressed and we say that the object is `screened'.  
\begin{figure}[!h]
\centerline{\epsfig{file=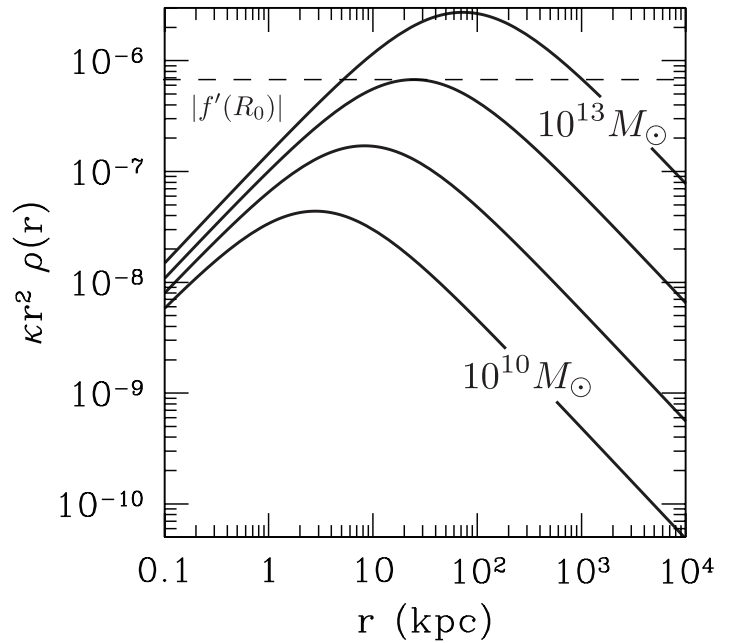,height=20pc,angle=0}}
\caption{The combination $\kappa \rho(r) r^2$ for an NFW profile [Eq.~\ref{eq:NFW}] for masses incremented by an order of magnitude between $10^{10}\ M_{\odot}$ (bottom curve) to $10^{13}\ M_{\odot}$ (top curve).  The outermost point where $|f'(R_0)|$ intersects these curves indicates the transition radius, $r_0$, inside of which deviations from relativity are suppressed.  For a given value of $|f'(R_0)|$ those halos with a mass below some threshold will not be screened and will therefore exhibit order unity modifications to gravity.}
\label{fig:NFW_halo}
\end{figure}

As shown in Refs.~\cite{Chiba:2006jp,Hu:2007nk} screening occurs within a radius $r_0$ implicitly given by
\begin{equation}
|f'(R_0)| < \kappa \rho(r_0) r_0^2,
\label{eq:non-linear}
\end{equation}
where $R_0$ is the value of the scalar curvature on cosmological scales and $\rho$ is the local value of the density.  Note that if the density of an object is too small then Eq.~(\ref{eq:non-linear}) is not satisfied at any radius and the object is completely unscreened with order unity deviations from GR throughout.  One can think of $|f'(R_0)|$ as determining a characteristic gravitational potential for the theory so that when the local gravitational potential ($\kappa \rho r^2$) is larger than this characteristic value deviations from GR are highly suppressed.  
Numerical solutions show that the transition from $R\sim R_0$ to $R_0 \ll R = - \kappa \rho$ occurs over a relatively short length-scale (see, e.g., Fig.~10 in Ref.~\cite{Hu:2007nk}) so we will approximate it by a step function.  

The most stringent observational constraint on $f(R)$ gravity theories comes from the requirement that it pass solar system tests.  Measurements of the motion of light in the solar system has placed the constraint $\gamma_{\rm PPN, \odot}=1+(2.1\pm 2.3)\times 10^{-5}$ \cite{Bertotti:2003rm,Shapiro:2004zz}.  In order for $f(R)$ gravity theories to pass solar system tests the theory must suppress deviations from GR within our halo leading to the constraint $|f'(R_0)| \lesssim 10^{-6}$ \cite{Hu:2007nk}.  

Measurements of lensing around other galaxies can also serve to constrain this theory.  In particular, Eq.~(\ref{eq:non-linear}) shows that $f(R)$-gravity predicts a lensing signal around galaxies which depends on halo mass.  To see this, consider an NFW halo \cite{Navarro:1996gj} of the form
\begin{equation}
\rho(r) = \rho_{\rm c} \delta_{\rm c} \left(\frac{r}{r_s}\right)^{-1}\left(1+\frac{r}{r_s}\right)^{-2}, 
\label{eq:NFW}
\end{equation}
where $r_s$ is the scale radius and $\rho_{\rm c}$ is the critical density of the universe.  The amplitude $\delta_{\rm c}= (\Delta/3) c^3/[\ln(1+c) - c/(1+c)]$ relates the concentration to the virial radius with an overdensity, $\Delta = 119$.  We also assume the mass-concentration relation
\begin{equation}
c = \frac{9}{1+z} \left(\frac{M}{8.12 \times 10^{12} h^{-1} M_{\odot}}\right)^{-0.14}, 
\end{equation}
where $h$ is the Hubble parameter in units of 100 km/(s Mpc) \cite{Bullock:1999he,Eke:2000av}.  This relation reduces the NFW profile to a one-parameter family which we take to be dependent on the virial mass, $M$.  Since $\rho_{\rm NFW} r^2 \propto r$ for $r < r_s$ and $\rho_{\rm NFW} r^2 \propto r^{-2}$ for $r > r_s$ it is clear that the innermost point at which deviations from GR are suppressed in $f(R)$-gravity will occur at the scale radius $r_s$.  

\begin{figure}[!h]
\centerline{\epsfig{file=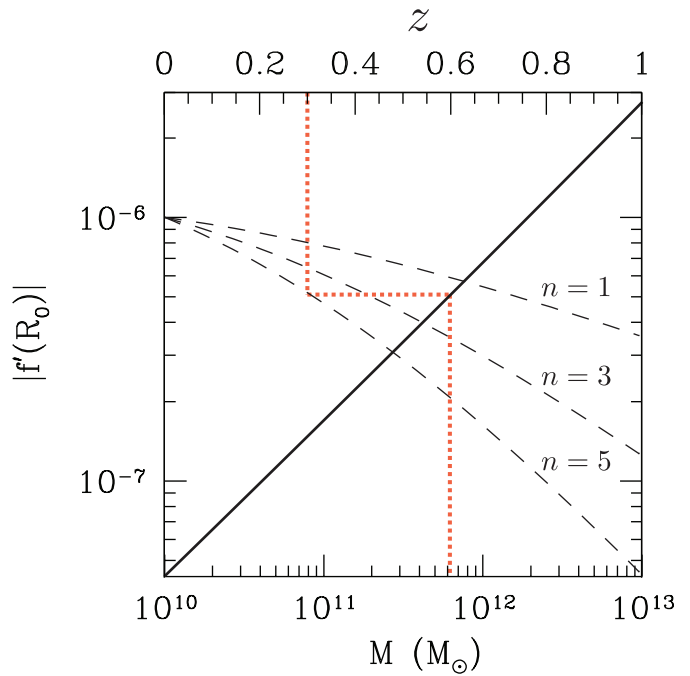,height=21pc,angle=0}}
\caption{The minimum mass for a halo to be screened given a value for $|f'(R_0)|$.  Solar system tests 
require that the Milky Way's halo be screened.  Since the halo has a mass $\sim 10^{12} \ M_{\odot}$ this
implies that $|f'(R_0)| \lesssim 10^{-6}$ \cite{Hu:2007nk}.  Strong lensing measurements around galaxies with smaller
masses will have $\gamma_{\rm PPN} = 1/2$.  The dashed line and upper $x$-axis show how $|f'(R_0)|$ varies as a function of redshift with $|f'(R_0)|(z=0) = 10^{-6}$.  In general, for $f(R)$ models that produce late-time acceleration $|f'(R_0)| \propto 1/H(z)^n$.  In this figure we show the evolution of $|f'(R_0)|$ for the model presented in Ref.~\cite{Hu:2007nk}.  The dotted red line indicates the way in which this plot should be read: at a given redshift (upper $x$-axis) a vertical line intersects a given $f(R)$ model (dashed curve); a horizontal line then intersects the solid `screening' line; from that point a vertical line drawn to the corresponding halo mass (lower $x$-axis) gives the mass above which the halo is screened at that redshift.}
\label{fig:mass_F_R}
\end{figure}

 Looking at Eq.~(\ref{eq:non-linear}) and the NFW density profile we can see that halos with masses which satisfy
 \begin{equation}
 |f'(R_0)| > 2 \kappa \rho_c \delta_c(M)
 \label{eq:fRcondition}
 \end{equation}
will not be screened\footnote{Strictly speaking this condition only applies to halos which are isolated. A strong lens galaxy which sits within a larger halo may be screened by the larger halo even though the mass of the lens halo is below the threshold given in Eq.~(\ref{eq:fRcondition}) \cite{Schmidt:2010jr}.  The local environments of strong lens galaxies can be approximately determined using photometric data and shows that for the data considered in this paper (the SLACS survey) the richness of the lens systems is on average a few with very few close companions \cite{2008MNRAS.383L..40A} indicating that most systems should be sufficiently isolated for our purposes here.}.  Since $\delta_c(M)$ decreases with decreasing $M$ this sets an upper limit to the 
mass of halos which can be screened given a value for $|f'(R_0)|$.  This dependence is shown
as the solid line in Fig.~\ref{fig:mass_F_R}.  
For masses below this threshold the theory deviates from GR by factors of order unity and strong lensing around these halos have 
$\gamma_{\rm PPN} = 1/2$.  Therefore, a measurement of $\gamma_{\rm PPN} = 1$ in lower mass galaxies 
would place a more stringent constraint on $|f'(R_0)|$. 

Finally, we must also take into account that $|f'(R_0)|$ depends on redshift.  For $f(R)$ models which produce late-time acceleration we have $|f'(R_0)| \propto 1/H^n$, where $n>0$ so that $|f'(R_0)|$ increases as the universe expands.  For a galaxy with $\rho r^2 \propto r^{-m}$ with $m>0$ (which will be true in the outer regions of the galaxy to ensure that the galaxy has a finite mass) this causes the $\gamma_{\rm PPN}=1/2$ region to propagate inwards as time progresses so that $f(R)$-gravity predicts a redshift dependent $\gamma_{\rm PPN}$ as well.  The dashed curves in Fig.~\ref{fig:mass_F_R} show the evolution of $|f'(R_0)|$ as a function of redshift for the specific $f(R)$ model found in Ref.~\cite{Hu:2007nk}, 
\begin{equation}
f(R) = - m^2 \frac{c_1 (R/m^2)^{n}}{c_2 (R/m^2)^{n+1}+1}, 
\label{eq:fR_model}
\end{equation}
where $m^2 \sim H_0^2$ and $n$ is a free index and the ratio $c_1/c_2$ is set by requiring the model have the same expansion history as in a $\Lambda$CDM model with $\tilde{\Omega}_M$ and $\tilde{\Omega}_{\Lambda}$
\begin{equation}
\frac{c_1}{c_2} \approx 6 \frac{\tilde{\Omega}_M}{\tilde{\Omega}_{\Lambda}}.  
\end{equation}
Therefore these models have two free parameters which we choose to be $n$ and $|f'(R_0)|(z=0)$. 

\section{Strong lensing parameterized by gravitational slip \label{sec:slip}}

In order to test the predictions of GR we must compare predictions and observations with other theories of gravity.  Besides looking at other theories on a case-by-case basis it is more useful to parameterize modifications to GR and constrain the value of those parameters.  This approach has proven very sucessful when interpreting observations of the effects of gravity in the solar system in the form of the parameterized post-Newtonian (PPN) formalism \cite{Will:1981cz}.  In this formalism the relationship between different parts of the metric are parameterized by constant coefficients and observations of the motion of test particles (both massive and massless) measure the values of these coefficients.  

The PPN formalism relies on the presence of localized sources of stress-energy and so cannot be applied without change to a cosmological context.  Some studies have attempted to articulate ways in which to extend the PPN formalism to cosmological observations \cite{Hu:2007pj,Caldwell:2007cw,Amin:2007wi,Bertschinger:2008zb}.  The basic idea of all of the currently proposed parameterizations is that current cosmological observations using the cosmic microwave background, weak lensing, and evolution of large-scale structure, are only sensitive to the scalar part of the metric [i.e., the two Newtonian potentials $\Psi$ and $\Phi$ in the metric given in Eq.~(\ref{eq:scalar_metric})].  Furthermore, since any modifications to GR that accounts for a phase of late-time accelerated expansion is significant for $z\lesssim 1$ the only perturbed fluid variables that are dynamically important are the matter density ($\delta_M$) and velocity perturbations ($\theta_M$).  Considering only those gravity theories where stress-energy is conserved gives two evolution equations (the continuity and Euler equations).  Therefore a generic modified gravity theory is defined when two gravitational field equations are specified. 

The two scalar potentials form a linear combination which is determined by a Poisson-like equation but with a time and space dependent Newton's `constant' $\mu(z,\vec{x})$
\begin{equation}
\frac{\nabla^2 \left(A \Phi + B \Psi \right)}{A+B} = 4 \pi \mu(z,\vec{x}) G \rho,
\end{equation}
and their ratio can be parameterized by another time and space dependent function given in Eq.~(\ref{normal_gamma1}) and repeated here, 
\begin{equation}
\frac{\Phi}{\Psi} = \gamma_{\rm PPN}(z,\vec{x}).
\end{equation}
In GR we have $A=1$, $B=0$, $G(z,\vec{x}) = G$, and $\gamma_{\rm PPN}(z,\vec{x}) = 1$.  Particular modified gravity theories can then be parameterized by how the functions $G(z,\vec{x})$ and $ \gamma_{\rm PPN}(z,\vec{x})$ depend on scale and time \cite{Hu:2007pj,Amin:2007wi}.  

One particular parameterization, first proposed in Ref.~\cite{Caldwell:2007cw}, supposes $\gamma_{\rm PPN}(z, \vec{x}) = [1+ \varpi_0' \rho_{\rm DE}/\rho_M(z)]^{-1} = [1+ \varpi_0/(1+z)^{3}]^{-1}$ so that order unity modifications turn on around the transition from matter domination to dark energy domination.  As described in more detail in Ref.~\cite{Daniel:2009kr} this parameterization further chooses to maintain the scalar part of the $(0,i)$ component of the Einstein field equations leading to a time dependent $\mu$.  In the next Section we show how observations of strong lensing around galaxies over a range of redshifts are able to constrain the value of $\varpi_0$. 

\section{$\gamma_{\rm PPN}$ from measurements of strong lenses \label{measurements}}

The original idea of measuring $\gamma_{\mathrm{PPN}}$ from observations of strong lenses was first discussed in Ref.~\citep{sirousse} and was first applied to data in Ref.~\citep{Bolton:2006yz}.  A qualitative understanding of how observations of strong lenses can yield a measurement of $\gamma_{\rm PPN}$ can be understood by considering the following simplified 
example \cite{sirousse}.  In this example the lens density distribution is given by a singular isothermal sphere with the observed line-of-sight velocity dispersion $\sigma_{\rm obs}$.  The lens then produces an Einstein ring with a radius, $R_E$, given by \cite{1992grle.book.....S}
\begin{equation}
R_{E} =4\pi \sigma_{\rm obs}^2 \left(\frac{1 + \gamma_{\rm PPN}}{2}\right)  \frac{D_L D_{LS}}{ D_S},
\end{equation}
where $D_X$ is the angular diameter distance to the lens ($L$), source ($S$), and between the lens and source ($LS$).
The observed spectra give measurements of the source and lens redshifts
as well as of the line-of-sight velocity dispersion \citep{1992MNRAS.254..389R}.  The angular diameter distances are obtained by fixing a fiducial cosmology although the choice of cosmological parameters does not significantly impact the final result.  The data then yields a measurement of
\begin{equation}
 \gamma_{\mathrm{PPN}} = 2 \pi \frac{D_S}{D_LD_{LS}} \frac{R_E c^2}{\sigma_{\rm obs}^2}-1.
 \end{equation}
As we will now describe, in practice the problem is more complicated since the density profile of the lens cannot be described by such a simple model.  
 
As in the simplified model that was just discussed, in order to measure $\gamma_{\rm PPN}$ we must relate the observed stellar velocity dispersion to the lensing observations.  First note that
in an analogy with Gauss' law the deflection angle in a modified gravity theory parameterized as in Eq.~(\ref{eq:mu}) and (\ref{normal_gamma1}) for a circularly symmetric lens depends on the enclosed mass as
\begin{equation}
\hat{\alpha} =  \frac{4  D_{LS}}{D_S} \frac{\mu G M(\theta)}{D_L\theta} \hat{\theta}, 
\end{equation}
where $\hat{\theta}$ is a unit vector projected on the sky centered at the lens, and $M(\theta)$ is the projected mass enclosed within the angle $\theta$.  
The lens equation \cite{1992grle.book.....S} then relates the observed Einstein radius $\theta_E = R_E/D_L$ to the enclosed mass and $\mu$
\begin{equation}
R_{E}^2 =4 \frac{D_L D_{LS}}{ D_S} \mu G M(R_E) .
\label{eq:SIS_E}
\end{equation}

It is useful to associate an effective `lensing' velocity dispersion with each lens system.  
The effective velocity dispersion, $\sigma_{\rm lens}$, is defined through the measured Einstein radius,
 \begin{equation}
R_E \equiv 4 \pi \frac{\sigma_{\rm lens}^2}{c^2} \frac{D_L D_{LS}}{D_S},
\label{einstein_radius}
\end{equation}
and the lens equation [Eq.~(\ref{eq:SIS_E})] allows us to relate this to the projected mass within the Einstein radius
\begin{equation}
M(R_E) = 4\pi^2 \frac{\sigma_ {\rm lens} ^4}{c^2 \mu G} \frac{D_L D_{LS}}{D_S}.
\label{einstein_mass}
\end{equation}
We can therefore write
\begin{equation}
\sigma_{\rm lens} ^2= \frac{1}{\pi} \frac{\mu G M(R_E)}{R_E} .
\end{equation}
This `lensing' velocity dispersion, $\sigma_{\rm lens}^2$, should not be confused with the \emph{observed} stellar velocity dispersion, $\sigma_{\rm obs}$. 
In order to measure $\gamma_{\rm PPN}$, the equations of hydrostatic equilibrium are used to relate the observed stellar velocity dispersion to $M(R_E)$. 

The analysis presented here extends the model due to Ref.~\citep{Koopmans:2005ig}.  The total mass density of the galaxy is modeled as a power law
\begin{equation}
\rho_M(r) = \rho_{M,0}\left( \frac{r}{r_*}\right)^{-p}.
\end{equation}
The stellar component is well fit by a Hernquist profile \cite{1990ApJ...356..359H}
\begin{equation}
\rho_{\rm L}(r) = \frac{M_* r_*}{2 \pi r(r+r_*)^3}, 
\end{equation}
where $M_*$  is the total stellar mass and $r_*$ is a scale radius which can be written in terms of the effective radius of an $R^{1/4}$ luminosity profile as $r_*= R_{\rm eff}/1.8153$.
Finally, we allow for a non-zero anisotropy in the stellar velocity ellipsoid which is constant with radius, 
\begin{equation}
\beta \equiv 1- \frac{\langle \sigma_{\theta}^2\rangle}{\langle \sigma_{r}^2\rangle}.
\end{equation}
The radial velocity dispersion is found by solving the spherically symmetric Jeans equation \citep{binney} for the stellar component.  Projecting out the line of sight velocity dispersion and performing a weighted average over the luminosity profile within a circular aperture of projected radius $R_A$ the observed stellar velocity dispersion is related to the model parameters through the expression
\begin{equation}
\sigma_{\rm obs}^2= \frac{1}{\pi}\frac{2 \mu G M_E(R_E)}{(1+\gamma_{\rm PPN})R_E} \left(\frac{R_E}{r_*}\right)^{p-2} g(p, \beta, R_A/r_*,\sigma_{\rm see}), 
\label{model_1}
\end{equation}
where $\sigma_{\rm see}$ \cite{1998gaas.book.....B} is the seeing (i.e., blurring due to atmospheric distortions), and $g(p, \beta, R_A/r_*,\sigma_{\rm see})$ can be written in terms of integrals over hypergeometric functions.  This 
 \begin{figure}[!h]
\centerline{\epsfig{file=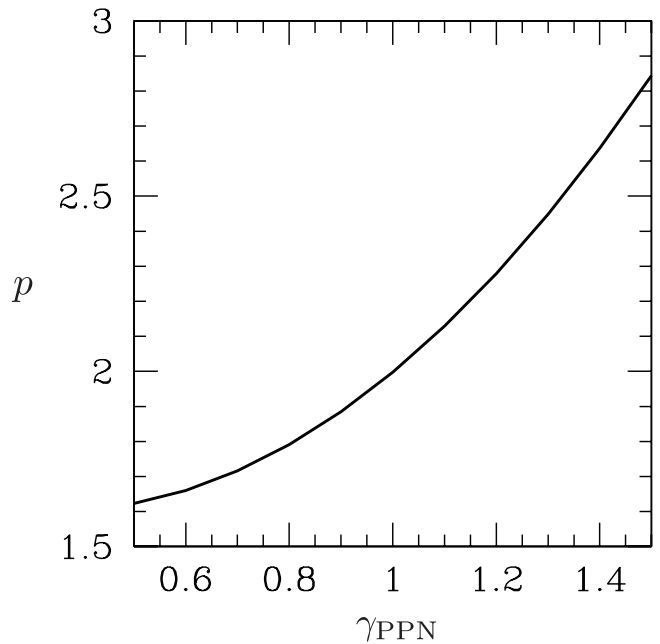,height=20pc,angle=0}}
\caption{The degeneracy between $\gamma_{\rm PPN}$ and the slope of the total matter density $p$.  In order to generate this curve we have used the mean values from the SLACS survey: $R_E = R_{\rm eff}/2 = 5$ kpc with $z_{\rm L} = 0.1$ and $\beta = 0$ .}
\label{fig:degen}
\end{figure}
\begin{figure*}[!ht]
\centerline{\epsfig{file=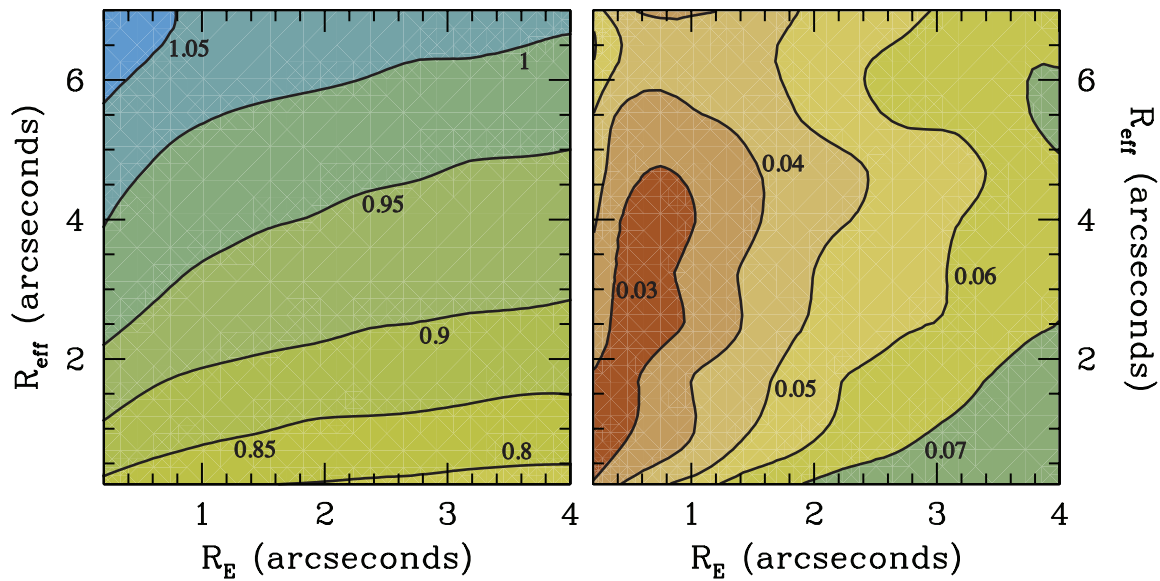,height=18pc,angle=0}}
\caption{The mean (\emph{left}) and dispersion (\emph{right}) for the distribution of $\sqrt{2}f_{\sigma}/\sqrt{1+\gamma_{\rm PPN}}$ as a function of the observed effective radius, $R_{\rm eff}$, and the observed Einstein radius, $R_E$.  The intrinsic distribution for the slope of the total density profile, $p$, and the velocity anisotropy, $\beta$, were assumed to be equal to their intrinsic distributions measured in early-type galaxies at low redshift where more detailed measurements of the stellar kinematics can be measured.}
\label{sigma_contour}
\end{figure*}

The ratio between the observed stellar velocity dispersion and the lens velocity dispersion, $f_{\sigma}^2 \equiv \sigma_{\rm obs}^2/\sigma_{\rm lens}^2$ is given by
\begin{eqnarray}
f_{\sigma}^2 &=& \left(\frac{2}{1+\gamma_{\rm PPN}}\right)\left(\frac{R_E}{r_*}\right)^{p-2} g(p, \beta, R_A/r_*,\sigma_{\rm see}).\nonumber \\ 
\label{eq:MC}
\end{eqnarray}
Observations of the stellar velocity dispersion and the Einstein radius yield measurements of $f_{\sigma}$.  However, from Fig.~\ref{fig:degen} it is clear there is a degeneracy between the slope of the density profile, $p$, and $\gamma_{\rm PPN}$.  In order to make progress in measuring $\gamma_{\rm PPN}$ a prior must be placed on $p$.  It is important that any approach taken to place this prior be independent of the theory of gravity.  One approach, discussed in Ref.~\cite{Koopmans:2009av}, uses an assumed scaling law (related to the fundamental plane \cite{2005ApJ...623..666R}) that relates the power-law slope, $p$, to the luminosity, Einstein radius, and effective radius is used to estimate the average value for $p$ within a population of lenses.  In this approach a universal value for $\gamma_{\rm PPN}$ would appear as a constant offset and does not affect the estimate of $\langle p \rangle = 1.959 \pm 0.077$ \cite{Koopmans:2009av}.  However, we are interested in exploring whether $\gamma_{\rm PPN}$ may be non-universal.  Furthermore, the analysis in Ref.~\cite{Koopmans:2009av} is done in the limit where the dark matter fraction of  in the system is 1.  This is a significant simplification since the average dark matter fraction within $R_E$ of a subset of these systems has been estimated to be 25\% \cite{2006ApJ...649..599K}.  

Another approach is to place a prior on $p$ given by its distribution measured in low redshift early-type galaxies where more detailed kinematic data can be used to determine the full density profile.  This is the approach taken here and gives $\langle p \rangle = 1.93$, $\sigma_{p} = 0.08$ and $\langle \beta \rangle = 0.18$ and $\sigma_{\beta} = 0.13$ \cite{2000A&AS..144...53K,2001AJ....121.1936G}.  But as is clear from Fig.~\ref{fig:degen}, the constraint on $\gamma_{\rm PPN}$ is very sensitive to the assumed $\langle p \rangle$ (see Fig.~\ref{fig:degen}).  To remove this dependence on $\langle p \rangle$ in our final constraints we marginalize over the average of value $\gamma_{\rm PPN}$ within the sample.  Therefore, our final constraints come from the lack of any significant correlation between $\gamma_{\rm PPN}$ and properties of the lens such as its redshift and mass. 

Using Eq.~(\ref{eq:MC}) the resulting distribution for $\sqrt{2}f_{\sigma}/\sqrt{1+\gamma_{\rm PPN}}$ was calculated on a grid of values for $0.2'' \leq R_{\rm eff}\leq 7''$ and $0.2''\leq R_E\leq 4''$ using an aperture radius $R_{A} = 1.5''$ and seeing $\sigma_{\rm see} = 0.64''$ \cite{2006ApJ...649..599K} and is shown in Fig.~\ref{sigma_contour}. 
For a value of $\gamma_{\rm PPN}$, an underlying probability distribution for $p$ and $\beta$, and measured values for $R_E$ and $R_{\rm eff}$, Eq.~(\ref{eq:MC}) gives the underlying probability distribution of $f_{\sigma}$, denoted by $(d P/d f_{\sigma})\big(\gamma_{\rm PPN}\big)$.  The likelihood for $\gamma_{\rm PPN}$ is then given by
\begin{eqnarray}
\mathcal{L} (\gamma_{\rm PPN}) &=& \\
&&\int 
\frac{d P}{d f_{\sigma}}\big(\gamma_{\rm PPN}\big)[f_{\sigma} | R_E, R_{\rm eff}] G(f_{\sigma}; \sigma_{f_{\sigma}}, f_{\sigma}^0) d f_{\sigma},\nonumber
\end{eqnarray}
where $G(f_{\sigma}; \sigma_{f_{\sigma}}, f_{\sigma}^0) $ is a Gaussian with a mean equal to the measured value of $f_{\sigma}^0$ and a standard deviation, $\sigma_{f_{\sigma}}$, equal to the observational error on $f_{\sigma}$.  

\begin{figure}[!ht]
\centerline{\epsfig{file=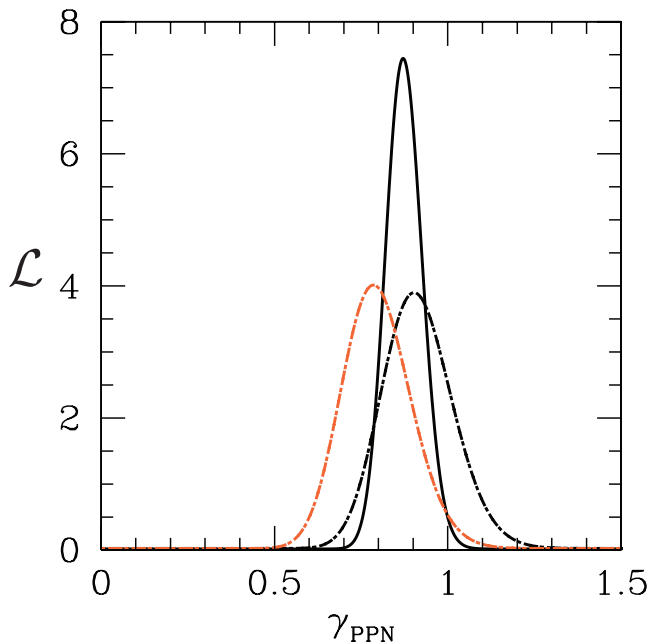,height=20pc,angle=0}}
\caption{The joint likelihood for a universal $\gamma_{\rm PPN}$.  The solid black curve is the joint likelihood for the full SLACS sample (53 systems) \cite{Bolton:2008xf} which gives $\gamma_{\rm PPN} = 0.88 \pm 0.05$; the black dashed-dot curve on the right is uses the original 15 systems considered in Ref.~\cite{Bolton:2006yz} which gives $\gamma_{\rm PPN} = 0.93 \pm 0.1$; the red dashed-dot curve on the left is a re-analysis of the 14 systems in common between the original 15 and the full 53 systems which gives $\gamma_{\rm PPN} = 0.81 \pm 0.1$.  As described in Ref.~\cite{Bolton:2008xf} the re-analysis of these 14 common systems reflects differences in the mass modeling, surface brightness measurements, and measurements of the velocity dispersion.  The effect these differences have on the inferred constraint to $\gamma_{\rm PPN}$ is well within 1$\sigma$ and so is well described by the errors we have included in the calculation of the likelihood (which are dominated by errors on the measured velocity dispersion). All quoted errors are at 68\% c.l.}
\label{universal_Like}
\end{figure}
Since the resulting probability distribution of $\sqrt{2}f_{\sigma}/\sqrt{1+\gamma_{\rm PPN}}$ for fixed $R_{\rm eff}$ and $R_E$ is well approximated by a Gaussian it is straight-forward to derive the distribution of $f_{\sigma}$ for any value of $\gamma_{\rm PPN}$, $R_{\rm eff}$, and $R_E$.

We first consider the case where $\gamma_{\rm PPN}$ is a universal constant and the results are shown in Fig.~\ref{universal_Like}.  In that figure the solid black curve shows the likelihood for the full SLACS survey (53 systems) \cite{Bolton:2008xf}.  The full joint likelihood gives the result $\gamma_{\rm PPN} = 0.88 \pm 0.05$ at 68\% c.l.  As we discuss in more detail below, even though this result seems to be in conflict with GR, it is more of a reflection of a difference between the mean slope of the mass density of the galaxies in the local universe as compared to the SLACS galaxies.  The other two dashed-dot curves are the likelihoods when considering a subset of the full survey.  The dot-dashed black curve on the right is the likelihood for the original 15 systems considered in Ref.~\cite{Bolton:2006yz} and the  dot-dashed red curve on the left uses 14 of those systems but reflects differences in the mass modeling, surface brightness measurements, and measurements of the velocity dispersion.   As discussed in Ref.~\cite{Bolton:2008xf} the difference in mass estimates, measured surface brightness and velocity dispersion, provides a more realistic sense of the measurement errors for these quantities. The dominant effect on the measurement of $\gamma_{\rm PPN}$ is the error in the measured stellar velocity dispersion. The analysis presented here was slightly more conservative than Ref.~\cite{Bolton:2008xf} and took the statistical errors in $\sigma_S$ quoted in Ref.~\cite{Bolton:2008xf} while enforcing a minimum of 7\%.  The agreement, at the 1$\sigma$ level, between the original analysis (dot-dashed black) and the modified analysis (dot-dashed red) indicates the error budget used in this work appropriately incorporates errors in the mass modeling, surface brightness measurements, and velocity dispersion. 

Comparing the constraints presented here to the results found in Ref.~\cite{Bolton:2006yz}, $\gamma_{\rm PPN} = 0.98 \pm 0.07$, we find a similar error but a significantly lower best fit value.  This difference can be explained by noting that an analysis of the SLACS lenses assuming GR (i.e., $\gamma_{\rm PPN} = 1$) found an intrinsic distribution for the slope of the matter density, $p$, of $\langle p \rangle  = 2.0$ and $\sigma_{p} = 0.12$ \cite{2006ApJ...649..599K,Koopmans:2009av}.  Since the mean of this distribution is approximately 0.1 away from the mean of the low-$z$ distribution used to constrain $\gamma_{\rm PPN}$ (i.e., $\langle p \rangle = 1.93$) and given that the degeneracy between $p$ and $\gamma_{\rm PPN}$ is nearly linear (see Fig.~\ref{fig:degen}) it follows that the best-fit value should be about 0.1 away from $\gamma_{\rm PPN}=1$.  The main difference between the analysis presented here and the one in Ref.~\cite{Bolton:2006yz} is that this analysis uses a more realistic model for the luminosity profile of the lens galaxy (a Hernquist profile), which was also used to measure $p$ in the original SLACS data \cite{2006ApJ...649..599K}, as opposed to Ref.~\cite{Bolton:2006yz} which used a power-law profile. 
\begin{figure}[!ht]
\centerline{\epsfig{file=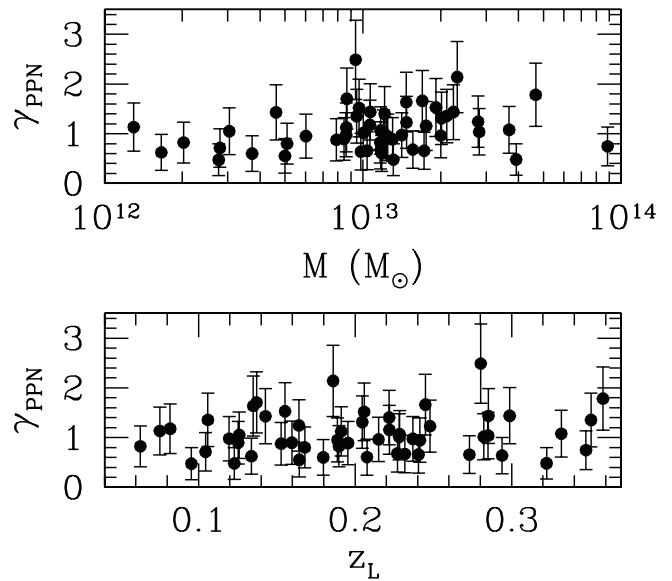,height=18pc,angle=0}}
\caption{The measurement of $\gamma_{\rm PPN}$ in individual lens systems as a function of the virial mass of the lens halo (top) and lens redshift (bottom).}
\label{gamma_plot}
\end{figure}

The degeneracy shown in Fig.~\ref{fig:degen} has a significant effect when applying the low redshift measurements of the slope of the total mass density $p$ to the SLACS objects and leads to an \emph{apparent} disagreement with GR at the 2.4$\sigma$ level and it is clear that this constraint on $\gamma_{\rm PPN}$ is of limited interest.  In particular, the lack of knowledge of the mean value of the slope of the total mass density in the lens galaxies leads to a bias in any constraint on a universal value of $\gamma_{\rm PPN}$.  In order to remove this bias we marginalize over the mean value of $\gamma_{\rm PPN}$ for the entire sample of lenses.  In this way the data does not constrain a universal (constant) value of $\gamma_{\rm PPN}$ but instead gives robust constraints on how $\gamma_{\rm PPN}$ may depend on properties of the lens such as its redshift and mass.  As we saw in previous sections, various theories of gravity give specific predictions on how $\gamma_{\rm PPN}$ depends on both lens redshift and mass.  

In order to demonstrate how a variable $\gamma_{\rm PPN}$ is constrained we applied the SLACS data to the two modified gravity theories described in Sec.~\ref{ST} and \ref{sec:fR}.  A variable $\gamma_{\rm PPN}$ is constrained by calculating the likelihood
\begin{equation}
\mathcal{L} \propto e^{- \chi^2/2}, 
\end{equation}
with
\begin{equation}
\chi^2= \sum_i \left\{\left[\gamma_{\rm PPN, i}^{\rm obs}- A \gamma_{\rm PPN}(z_{Li}, M_i)\right]/\sigma_{\gamma_{\rm PPN, i}}\right\}^2,
\end{equation}
where $\gamma_{\rm PPN, i}^{\rm obs}$ is the observed value, $\sigma_{\gamma_{\rm PPN, i}}$ is the error on the observation, $\gamma_{\rm PPN}(z_{L,i}, M_i)$ is the predicted value which depends on either the lens redshift or its mass, and $A$ is an amplitude which takes values between 0.8 and 1.2.  Marginalizing over $A$ (which is similar to marginalizing over the bias when using measurements of large-scale galaxy clustering) allows us remove the degeneracy between $\langle p \rangle$ and $\gamma_{\rm PPN}$ (which, as discussed before, is nearly linear; see Fig.~\ref{fig:degen}) and to explore constraints which arise due to (the lack of) any correlation or observed relationship between $\gamma_{\rm PPN}$ and lens redshift or halo mass. 

The lens redshift is a directly observable quantity whereas the mass is not.  To estimate the total virial mass, $M$, from observations we use an approximate relation with the observed half-light radius $R_{\rm eff}$.  Using the fact that the stellar to virial mass ratio is approximately 0.01 and given the relation between the stellar mass and effective radius found in Ref.~\cite{Shen:2003sda} for elliptical galaxies we can write
\begin{equation}
M = 7.5\times10^{11} \left(\frac{R_{\rm eff}}{{\rm kpc}}\right)^{1.78}\ M_{\odot}.
\end{equation}
Fig.~\ref{gamma_plot} shows the measured value of $\gamma_{\rm PPN}$ as a function of both halo mass (top panel) and lens redshift (bottom panel). Note that the mean estimated virial mass is $11 \times 10^{12}\ M_{\odot}$ which compares well with the mean virial mass determined through weak lensing $\langle M \rangle = 14^{+6}_{-5} \times 10^{12}\ M_{\odot}$ which was made using a subset of the full 53 systems considered here \cite{Gavazzi:2007vw}.  

We may approximate a generic non-universal $\gamma_{\rm PPN}$ as depending linearly on some property of the lens.  In particular, letting $x$ denote a property of the lens (such as its redshift) we model this generic dependence as
\begin{equation}
\gamma_{\rm PPN}(x) = \gamma_0(m_x) + m_x \frac{x}{\Delta x}, 
\end{equation}
where $\gamma_0(m_x)$ is a constant which depends on the slope $m_x$ and $\Delta x$ is the range of $x$ over which we have observations (i.e., the range of lens redshifts for a given survey).  In order to remove any sensitivity to the mean value of $\gamma_{\rm PPN}$ we define
\begin{equation}
\gamma_0(m_x) \equiv \bar{\gamma}_{\rm PPN} - \frac{m_x \bar{x}}{\Delta x}, 
\end{equation}
where $\bar{\gamma}_{\rm PPN}$ is the mean value of $\gamma_{\rm PPN}$ for all of the lenses in the survey. 
Here we will only be interested on constraining how $\gamma_{\rm PPN}$ may depend on the lens mass and redshift.  In these cases, at the 68\% confidence level (c.l.), we find that $m_z = 0.15 \pm 0.24$ ($0.06 < z < 0.36$) and $m_M = 0.13 \pm 0.36$ ($10^{12} \lesssim M/M_{\odot} \lesssim 10^{14}$). 

\subsection{Constraints to scalar tensor gravity}

\begin{figure*}[!ht]
\centerline{\epsfig{file=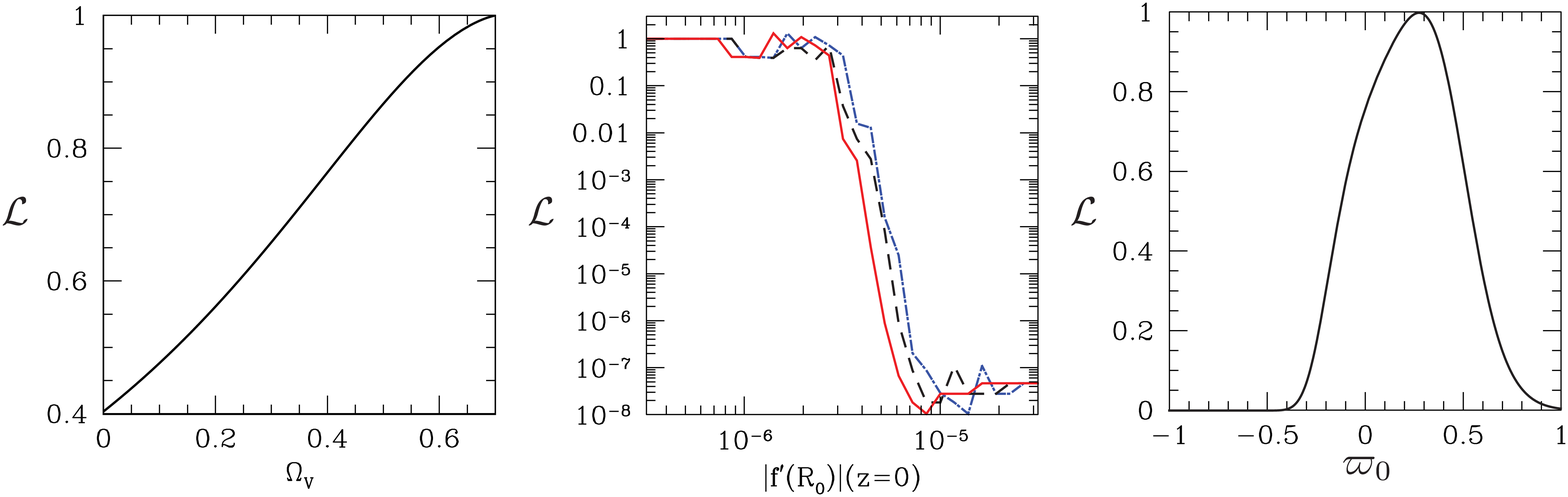,height=13.5pc,angle=0}}
\caption{Constraints to modified gravity theories and parameterizations using the SLACS data.  \emph{Left}: The left panel shows constraints to the energy density in a cosmological constant in units of the critical energy density, $\Omega_V$, in scalar tensor theories.  As discussed in Sec.~\ref{ST} scalar tensor gravity can explain the observed accelerated expansion with a value for the cosmological constant which is less than the value required in GR $\Omega_{\Lambda}$.  As $\Omega_V$ is made smaller the value of $\gamma_{\rm PPN}$ varies more with redshift leading to tension with the SLACS data (see Fig.~\ref{fig:transition}).  The SLACS measurements place the constraint $\Omega_V > 0.3$ at 68\% c.l. and $\Omega_V > 0.06$ at 95\% c.l.  \emph{Center}: The center panel shows constraints to the value of $|f'(R_0)|$ using the model presented in Eq.~(\ref{eq:fR_model}).  The three lines correspond to $n=1$ (solid red) $|f'(R_0)| (z=0) \lesssim 1.8 \times 10^{-6}$ 68\% ($2.5 \times 10^{-6}$ 95\%); $n=3$ (dashed black) $|f'(R_0)| (z=0) \lesssim 2 \times 10^{-6}$ 68\% ($2.8 \times 10^{-6}$ 95\%); $n=5$ (dot-dahsed blue) $|f'(R_0)| (z=0) \lesssim 2.2 \times 10^{-6}$ 68\% ($3 \times 10^{-6}$ 95\%).  \emph{Right}: The right panel shows constraints to the `gravitational slip' $\varpi_0$ which parameterizes the evolution of $\gamma_{\rm PPN}$ in time as discussed in Sec.~\ref{sec:slip}.  The SLACS data places the constraints $\varpi_0= 0.25^{+ 0.22}_{-0.27}\ ^{+0.45}_{-0.48}$ (68\%, 95\% c.l.) which is as restrictive as constraints derived from the cosmic microwave background, weak lensing, and evolution of large-scale structure \cite{Daniel:2009kr}.}
\label{constraints}
\end{figure*}

As described in Sec.~\ref{ST} scalar-tensor gravity generically predicts a redshift dependent $\gamma_{\rm PPN}$.  For the particular case where we account for the observed accelerated expansion using a scalar-tensor theory with a cosmological constant $\Omega_V < \Omega_{\Lambda}$ the gravitational lensing observations place constraints on the value of $\Omega_V$ as shown in the left panel of Fig.~\ref{constraints}.  
 As $\Omega_V$ is made smaller the value of $\gamma_{\rm PPN}$ varies more with redshift leading to tension with the SLACS data (see Fig.~\ref{fig:transition})
The data places the constraint $\Omega_V > 0.3$ at 68\% c.l. and $\Omega_V > 0.06$ at 95\% c.l.  Note that since we have marginalized over the ensemble average for $\gamma_{\rm PPN}$ these constraints rely solely on the lack of any significant correlation between $\gamma_{\rm PPN}$ and the lens redshift. 

Since these models are constructed to have negligible deviations from GR today measurements of either the expansion history or lensing and dynamics at high $z$ provide the only data which can place meaningful constrains on these theories. 

\subsection{Constraints to $f(R)$ gravity}

As described in Sec.~\ref{sec:fR} $f(R)$ gravity generically predicts a lensing signal which depends on the mass of lens halo.  Using the relationship between the mass threshold (below which $\gamma_{\rm PPN} =1/2$ and above which is equal to unity) shown in Fig.~\ref{fig:mass_F_R} the SLACS data places a constraint on the allowed values of $|f'(R_0)|(z=0)$.  Constraints to the particular model described in Eq.~(\ref{eq:fR_model}) are shown in the center panel in Fig.~\ref{constraints}.  Placing the prior $|f'(R_0)|(z=0) \leq 10^{-5}$ for the particular case where $n=1$ we find $|f'(R_0)| (z=0) \lesssim 1.8 \times 10^{-6}$ 68\% ($2.5 \times 10^{-6}$ 95\%); $n=3$ yields $|f'(R_0)| (z=0) \lesssim 2 \times 10^{-6}$ 68\% ($2.8 \times 10^{-6}$ 95\%); $n=5$ yields $|f'(R_0)| (z=0) \lesssim 2.2 \times 10^{-6}$ 68\% ($3 \times 10^{-6}$ 95\%).  The constraint to $|f'(R_0)|(z=0)$ is weakened as $n$ gets larger since, as demonstrated in Fig.~\ref{fig:mass_F_R}, the larger $n$ is the more $|f'(R_0)|(z)$ decreases with increasing redshift.  Therefore, for the same value of $|f'(R_0)|(z=0)$ a model with a larger $n$ will have a smaller value of $|f'(R_0)|(z\sim 0.2)$ leading to a weaker constraint at the SLACS redshifts.  

The constraints to $f(R)$ gravity from the SLACS data are about a factor of 10 worse than constraints from solar system tests.  However, they are several orders of magnitude better than constraints using observations of anisotropies in the CMB and measurements of the matter power spectrum \cite{Song:2006ej}.  A survey which could measure lensing and dynamics around low-mass galaxies ($M \lesssim 10^{12}\ M_{\odot}$) could potentially place constraints on $f(R)$ gravity which would improve upon solar system tests. 

Note that since we have marginalized over the mean value of $\gamma_{\rm PPN}$ these constraints rely solely on the lack of any significant correlation between $\gamma_{\rm PPN}$ and the lens mass. 
\vspace{-10pt}

\subsection{Constraints to the gravitational slip}

As described in Sec.~\ref{sec:slip} a particular way to parameterize deviations from GR is to introduce two new time and space dependent functions: a modified Newton's `constant' $\mu(z,\vec{x})$ and $\gamma_{\rm PPN}(z,\vec{x})$.  Assuming that the modifications of GR become important at the same time as the expansion starts to accelerate inspires the parameterization \cite{Caldwell:2007cw}
\begin{equation}
\gamma_{\rm PPN}(z) = \frac{1}{1+ \varpi_0(1+z)^{-3}}, 
\end{equation}
and measurements of the SLACS lenses places a constraint on $\varpi_0$ as seen in the right-most panel in Fig.~\ref{constraints}.  The SLACS data places the constraints $\varpi_0= 0.25^{+ 0.22}_{-0.27}\ ^{+0.45}_{-0.48}$ (68\%, 95\% c.l.) which is as restrictive as constraints derived from the cosmic microwave background, weak lensing, and evolution of large-scale structure \cite{Daniel:2009kr}.

Note that, as in the case of $f(R)$ gravity, since we have marginalized over the mean value of $\gamma_{\rm PPN}$ these constraints rely solely on the lack of any significant correlation between $\gamma_{\rm PPN}$ and the lens redshift. 

\section{Conclusions \label{conclusions}}
Constraints to modifications of GR from data taken within the solar system or from binary neutron star systems within our galaxy are very precise.  For instance, radar ranging to the Cassini spacecraft leads to a constraint $\gamma_{\rm PPN, \odot} = 1+(2.1\pm 2.3)\times 10^{-5}$ \cite{Bertotti:2003rm}.  Although these measurements place important constraints recent interest in modified gravity theories which can account for the observed accelerated expansion have focused attention on models with modifications that evolve with both time and environment.  

In the case of scalar-tensor theories of gravity time evolution is a natural consequence of introducing a new scalar degree of freedom.  In the case of $f(R)$ gravity the chameleon mechanism, where modifications to GR are suppressed in regions of high mass density, is a natural consequence of the fourth order nature of the modified field equations.  Other theories, such as DGP gravity \cite{Dvali:2000hr, Deffayet:2001pu} and the recently proposed Galileon \cite{Chow:2009fm}, also predict a non-universal $\gamma_{\rm PPN}$.  It is therefore important to not only look for ways to constrain $\gamma_{\rm PPN}$ in the local universe but to also investigate whether it may change depending on time, scale, mass, local environment, and so forth. 

Using measurements of stellar velocity dispersions and strong lensing around early-type galaxies from the full SLACS survey we have presented constraints to alternative gravity theories which can account for the observed late-time acceleration.  This analysis updates the results presented in Ref.~\cite{Bolton:2006yz} by including more realistic modeling of the stellar component as well as by using the full 53 systems in the SLACS survey.  We also extended the analysis beyond constraining a universal value for $\gamma_{\rm PPN}$ and applied the data to constrain $\gamma_{\rm PPN}$'s dependence on the mass and redshift of the lens.  

Constraints to a universal value for $\gamma_{\rm PPN}$ must be used with caution given the significant degeneracy between the slope of the total matter density, $p$, and $\gamma_{\rm PPN}$.  The two methods discussed in the text which estimate $\langle p \rangle$ independently of the theory of gravity both have unquantified systematic errors.  In the case where a scaling law is assumed the scaling law itself may introduce biases; the application of low-redshift observations to the higher redshift SLACS lenses may not be appropriate given that the structure of the galaxies may significantly evolve with redshift.  To remove this uncertainty the analysis presented here marginalized over the mean of the sample leading to constraints which only depend on how $\gamma_{\rm PPN}$ correlates with lens redshift and mass.  

Attempts to constrain modifications to GR on Mpc scales using observations of galaxy clusters, weak lensing, and galaxy surveys are complementary to the presented here \cite{White:2001kt,Schimd:2004nq,Jain:2007yk,Song:2008xd,Song:2008vm}.  Observations of galaxy clusters allow a measurement of $\gamma_{\rm PPN}$ though a comparison between the X-ray temperature or virial mass and measurements of strong lensing.  However, these observations have a limited statistical significance leading to constraints on $\gamma_{\rm PPN}$ to $\sim 50\%$ \cite{1988dama.conf..339N}.  This is partly due to the lack of large homogeneous samples of clusters.  It is also related to the fact that cluster dynamics are harder to model leading to larger systematic errors.  

Although the SLACS survey presents us with an opportunity to constrain the non-universality of $\gamma_{\rm PPN}$ the dynamical range of the the SLACS survey are limited: their redshifts range $0.05 \leq z \leq 0.35$ and their masses are of order $10^{13}\ M_{\odot}$.  Future surveys may be able to extend this to a higher redshift as well as to lower mass galaxies \cite{2009AAS...21332204M} which would improve the results presented here.  For instance, measurements of lensing around galaxies with masses $M < 10^{12}\ M_{\odot}$ could potentially place a more stringent constraint on $f(R)$ gravity theories than solar system tests.  Of course the challenge to measuring strong lensing around less massive galaxies is that the lensing cross section decreases with decreasing mass.  

\acknowledgments

After completing this paper, we became aware of similar work by Schwab, Bolton, and  Rappaport.  The author thanks Adam Bolton, Robert Caldwell, Daniel Grin, and Eric Linder for a careful reading of a previous version of the manuscript, and Kevin Bundy, Bhuvnesh Jain and Marc Kamionkowski for useful conversations.  This work was supported by the Berkeley Center for Cosmological Physics.  The author gratefully acknowledges the hospitality of the Aspen Center for Physics where some of this work was completed. 

\begin{appendix}
\section{Lensing and dynamics in weak-field limit of modified gravity theories \label{basic lensing app}}
We start with the line element corresponding to weak gravity, 
\begin{equation}
ds^2 = -(1+2 \Psi) dt^2 + (1-2 \Phi) \delta_{ij}dx^i dx^j.
\end{equation}
Einstein gravity is determined through the field equations
\begin{equation}
G_{\mu \nu} = \kappa T_{\mu \nu}, 
\end{equation}
where, writing the metric as $g_{\mu \nu} = \eta_{\mu \nu} + h_{\mu \nu}$  we have
\begin{equation}
G_{\mu \nu} = -\frac{1}{2}\Big[\Box \gamma_{\mu \nu} - \partial_{\nu} \sigma_{\mu} - \partial_{\mu} \sigma_{\nu} + \eta_{\mu \nu} \partial_{\rho} \sigma^{\rho}\Big],
\end{equation}
where $\gamma_{\mu \nu} \equiv h_{\mu \nu} - (1/2) \eta_{\mu \nu} h$, $h \equiv h_{\mu \nu}\eta^{\mu \nu}$, and $\sigma_{\mu} \equiv \partial^{\tau}\gamma_{\mu \tau}$.  Considering a scalar modification of the field equations to linear order (appropriate for several modified gravity theories that lead to a late-time accelerated expansion \cite{Boisseau:2000pr, EspositoFarese:2000ij,Perivolaropoulos:2005yv,Gannouji:2006jm,Carroll:2003wy,Capozziello:2003tk,Dvali:2000hr, Deffayet:2001pu, Chow:2009fm}) we write $G_{\mu \nu} \rightarrow \mu^{-1}G_{\mu \nu} + \eta_{\mu \nu} \alpha_1 + \alpha_2 \partial_{\mu} \partial_{\nu} \alpha_3 + \partial_{(\mu} \alpha_5 \partial_{\nu)} \alpha_6$, where $\alpha_2$ and $\alpha_4$ are evaluated on the background which may be time-dependent.  Now we specialize to a gauge in which we set
\begin{equation}
\sigma_{\mu} + \partial_{\mu} F = 0,
\end{equation}
with $F$ satisfying the equation
\begin{equation}
\partial_{\mu} \partial_{\nu} F= \alpha_2 \partial_{\mu} \partial_{\nu} \alpha_3 + \partial_{(\mu} \alpha_5 \partial_{\nu)} \alpha_6.
\end{equation}
With this gauge condition we can put the linearized field equation in the form
\begin{eqnarray}
\Box \gamma_{\mu \nu} &=& -2\kappa \mu \left(  T_{\mu \nu} -\frac{1}{2} T_{\rm eff} \eta_{\mu \nu}\right),
\end{eqnarray}
where we have defined
\begin{equation}
T_{\rm eff} \equiv \frac{2 \alpha_1 + \alpha_2 \Box \alpha_3}{\kappa},
\label{eq:Teff}
\end{equation}
and we can neglect $\partial^{\rho} \alpha_5 \partial_{\rho} \alpha_6$ since it will be second order in the perturbation. 
 We take the stress energy tensor to be dominated by pressure-less matter so that
\begin{equation}
T_{\mu \nu}(x) = \rho(x) \delta_{\mu 0} \delta_{\nu 0}.
\end{equation}
Specializing to a static source the solution to the field equation which is asymptotically flat becomes
\begin{equation}
\gamma_{\mu \nu} = -4 \mu \left\{\Phi_N \delta_{\mu 0} \delta_{\nu 0} +\eta_{\mu \nu} \Phi_{\rm eff}/2\right\},
\end{equation}
 with the `bare' potential given by 
 \begin{equation}
 \Phi_N(\vec{x}) = -\frac{\kappa}{8 \pi} \int \frac{\rho(\vec{x}')}{|\vec{x}-\vec{x}'|} d^3 x'
  \end{equation}
  and 
 \begin{equation}
\Phi_{\rm eff} (\vec{x}) = \frac{\kappa}{8\pi} \int \frac{T_{\rm eff}}{|\vec{x}-\vec{x}'|} d^3 x'.
\label{eq:phi_eff}
 \end{equation}
This new scalar degree of freedom will be determined through a field equation whose solution will depend on boundary conditions.  If we suppose that the modification introduces a new mass scale, $m$, then the effective potential may depend on $\Phi_{\rm eff}(\Phi_N, m r, z,\Omega_M, \dots)$.  
 
 Taking the line-element written in Eq.~(1) we have
 \begin{equation}
 h = 2\mu(\Psi-3 \Phi),
 \end{equation}
 so that
 \begin{eqnarray}
 \gamma_{00} &=& -\mu(3 \Phi + \Psi),\\
 \gamma_{ij} &=& \mu(\Phi-\Psi) \delta_{ij}.
 \end{eqnarray}
 We then have
 \begin{eqnarray}
 \Psi &=&\mu\left(\Phi_N + \Phi_{\rm eff}\right), \\
  \Phi &=& \mu \left(\Phi_N - \Phi_{\rm eff}\right).
 \end{eqnarray}
 
 \end{appendix}

\bibliography{bibliography}{}
\bibliographystyle{apsrev.bst}
\end{document}